# Determination of the thickness and orientation of few-layer tungsten ditelluride using polarized Raman spectroscopy

Minjung Kim[1], Songhee Han[1], Jung Hwa Kim[2], Jae-Ung Lee[1], Zonghoon Lee[2] and Hyeonsik Cheong[1,*]

[1] Department of Physics, Sogang University, Seoul 04107, Korea

[2] School of Materials Science and Engineering, Ulsan National Institute of Science and Technology, Ulsan 44919, Korea

E-mail: hcheong@sogang.ac.kr

**Keywords:** Tungsten ditelluride (WTe$_2$), Raman spectroscopy, polarization, orientation, thickness

**Abstract.** Orthorhombic tungsten ditelluride (WTe$_2$), with a distorted 1T structure, exhibits a large magnetoresistance that depends on the orientation, and its electrical characteristics changes from semimetallic to insulating as the thickness decreases. Through polarized Raman spectroscopy in combination with transmission electron diffraction, we establish a reliable method to determine the thickness and crystallographic orientation of few-layer WTe$_2$. The Raman spectrum shows a pronounced dependence on the polarization of the excitation laser. We found that the separation between two Raman peaks at ~90 cm$^{-1}$ and at 80−86 cm$^{-1}$, depending on thickness, is a reliable fingerprint for determination of the thickness. For determination of the crystallographic orientation, the polarization dependence of the $A_1$ modes, measured with the 632.8-nm excitation, turns out to be the most reliable. We also discovered that the polarization behaviors of some of the Raman peaks depend on the excitation wavelength as well as thickness, indicating a close interplay between the band structure and anisotropic Raman scattering cross section.

1. **Introduction**

Tungsten ditelluride (WTe$_2$), a layered 2-dimensional material in the transition metal dichalcogenide (TMD) family, has attracted much attention since a large, non-saturating magnetoresistance (MR) was reported [1–9]. Unlike other TMD materials such as MoS$_2$ which are semiconductors, WTe$_2$ is a semimetal in bulk and becomes insulating in the few-layer limit [8]. Furthermore, because WTe$_2$ has an orthorhombic, distorted-1T structure, its physical properties show pronounced anisotropy in the plane of the layers. For example, the MR behavior shows a strong anisotropy: when the magnetic field is applied to the *c* axis, MR is maximum when the current is parallel to the *a* axis and minimum along the *b* axis [1,3,5,6]. For these reasons, it is important to determine the thickness and the crystallographic orientation of WTe$_2$ when this material is used in MR devices, etc.

Polarized Raman spectroscopy is a powerful method for determining a number of layers and crystallographic orientation of layered 2-dimensional materials [2,8,10–27]. Several Raman studies on few-layer WTe$_2$ have been reported by several groups to correlated the Raman spectrum with the number of layers [8,23,26–28]. However, as will be shown below, some of the peaks in the spectrum are convolutions of two peaks and the line shape changes depending on the polarization. If this polarization dependence is not considered, determination of the layer thickness based on the position of the Raman peaks can be inaccurate. Polarized Raman spectroscopy was carried out on bulk WTe$_2$ to assign Raman modes [2], but no such work is reported yet for few-layer WTe$_2$. In this work, we investigated the polarization dependence of the Raman spectrum as a function of the thickness of WTe$_2$ and the excitation wavelength. We propose a reliable method to determine the number of layers and the crystallographic orientation of WTe$_2$ using polarized Raman spectroscopy.

## 2. Results and Discussion

*2.1. Determination of crystallographic orientation*

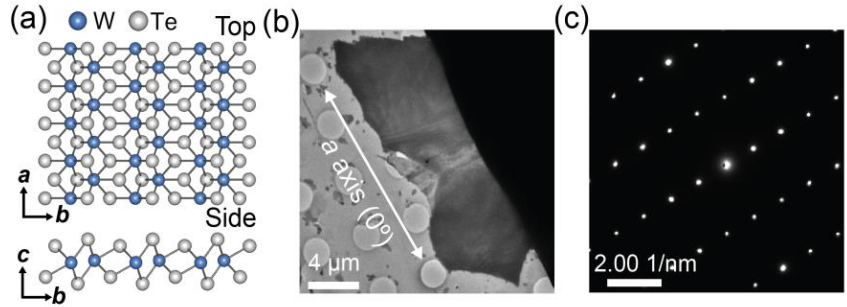

**Figure 1.** Crystallographic orientation of bulk WTe$_2$ sample. (a) Crystal structure of WTe$_2$. (b) Low magnification TEM image and (c) TED pattern of a bulk WTe$_2$ sample.

Figure 1(a) shows the crystal structure of WTe$_2$ with orthorhombic, distorted-1T structure with space group Pmn2$_1$ [2,21,23,26,28,29]. The *a*-axis is defined as the zigzag direction of the W-W chains. Figure 1(b) is a low magnification transmission electron microscope (TEM) image of a thick (bulk) WTe$_2$ sample on TEM grid, and Figure 1(c) is a corresponding transmission electron diffraction (TED) pattern. The *a*-axis direction determined from the TED pattern is indicated in Figure 1(b). Unlike the case of black phosphorus, another anisotropic 2-dimensional material, long straight edges do not necessarily correspond to a principal axis (Supplementary Figure 1).

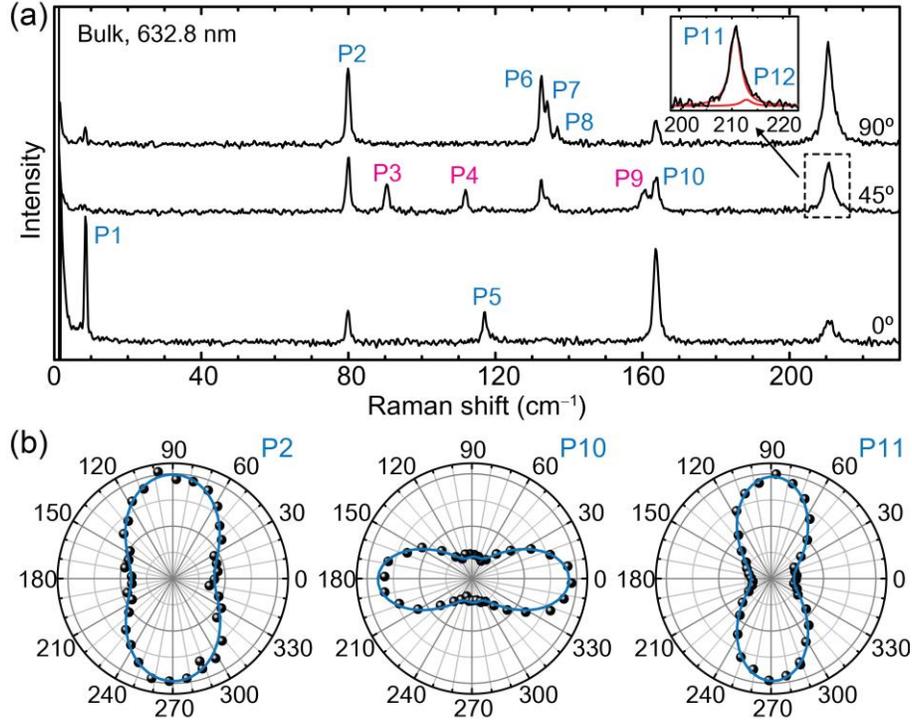

**Figure 2.** Polarization dependence of Raman spectrum of bulk $WTe_2$ sample. (a) Polarized Raman spectra of bulk $WTe_2$ in parallel polarization. Polarization angles with respect to the *a*-axis are indicated. The excitation wavelength is 632.8 nm. (b) Polarization dependence of intensities of peaks P2, P10, and P11. The radial scale of all polar plots in this paper starts at zero.

Figure 2(a) shows polarized Raman spectra of this sample in parallel polarization, using the excitation wavelength of 632.8 nm. The polarization angles are measured with respect to the *a*-axis direction determined from the TED measurement. The spectrum varies significantly depending on the polarization, and so the spectra taken in three representative polarizations are shown. For convenience, those peaks that are clearly resolved are numbered from the lowest frequency peak (P1). The optical phonon modes of space group $Pmn2_1$ are $A_1$, $A_2$, $B_1$, and $B_2$ [2,21,23,26,28–30], but $B_1$ and $B_2$ modes are forbidden in the back-scattering geometry. The complex Raman tensors, which take absorption into account [31–35], of $A_1$ and $A_2$ modes can be written as

$$A_1 = \begin{pmatrix} |a|e^{i\phi_a} & 0 & 0 \\ 0 & |b|e^{i\phi_b} & 0 \\ 0 & 0 & |c|e^{i\phi_c} \end{pmatrix}, \quad A_2 = \begin{pmatrix} 0 & |d|e^{i\phi_d} & 0 \\ |d|e^{i\phi_d} & 0 & 0 \\ 0 & 0 & 0 \end{pmatrix}. \tag{1}$$

The Raman intensity in parallel polarization is given by

$$I(A_1) \propto |a|^2 \cos^4\theta + |b|^2 \sin^4\theta + 2|a||b|\cos^2\theta \sin^2\theta \cos\phi_{ab}, \tag{2}$$

and

$$I(A_2) \propto |d|^2 \cos^2\theta \sin^2\theta, \tag{3}$$

where $\phi_{ab} = \phi_a - \phi_b$, and the angle $\theta$ is measured with respect to the $a$-axis. By comparing with Eqs. (2) and (3), we can assign P1, P2, P5, P6, P7, P8, P10, P11, and P12 as $A_1$ modes, and P3, P4, and P9 as $A_2$ modes (Supplementary Figure 2), which is consistent with theoretical predictions [2,21,23,26,28]. The atomic displacements of some of these modes can be found in Refs. [21,23,26]. Figure 2(b) shows that the polarization dependence of the $A_1$ modes can be used to determine the crystallographic orientation of the sample: the polarization direction in which the intensity of P2 is maximum corresponds to the $b$-axis.

## 2.2. Determination of thickness

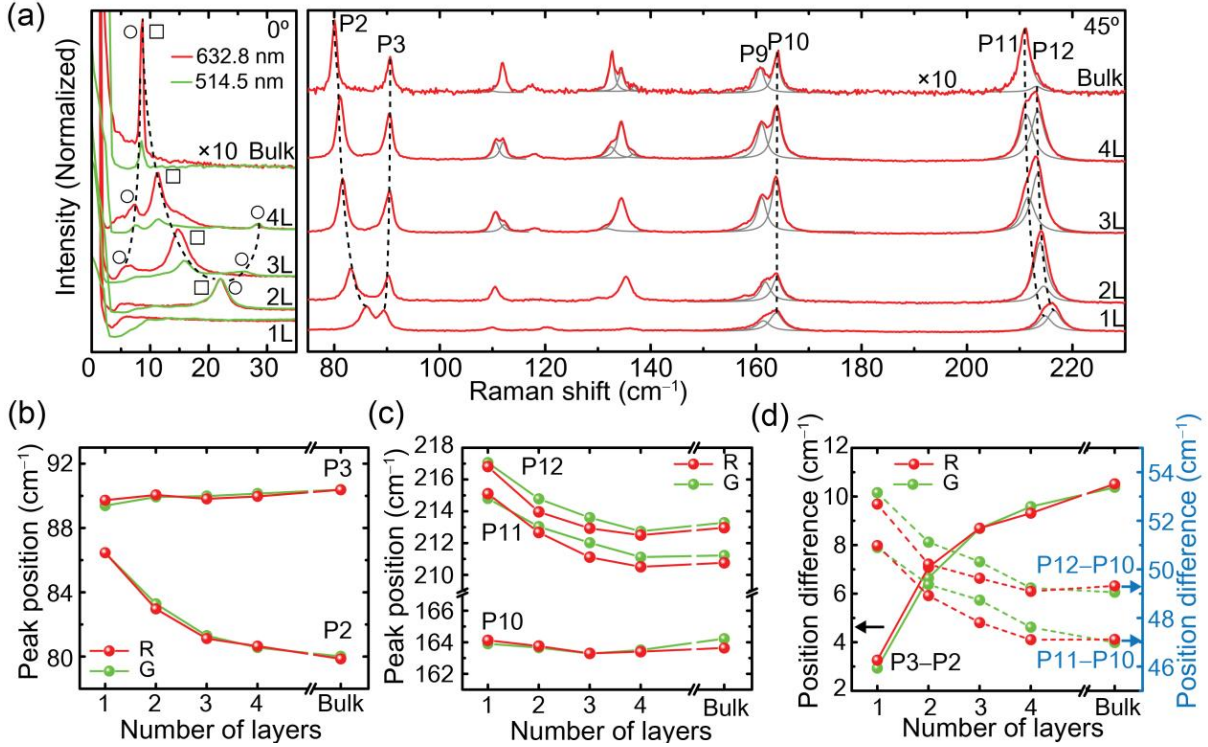

**Figure 3.** Thickness dependence of Raman spectrum of $WTe_2$. (a) Raman spectrum as a function of thickness of $WTe_2$, measured with excitation wavelength of 632.8-nm. For the low-frequency region, the spectra measured with 514.5-nm excitation are also shown. Shear (○) and breathing (□) modes are indicated. The Raman spectrum of bulk $WTe_2$ is magnified 10 times. Thickness dependence of (b) P2 and P3 and (c) P10, P11, and P12 peak positions. (d) Thickness dependence of peak position differences between P2 and P3; P10 and P11; and P10 and P12.

Figure 3(a) shows the Raman spectra as a function of thickness. For the low-frequency range (< 35 cm$^{-1}$), the Raman spectra for the polarization angle of 0° are shown because the low-frequency peaks have the maximum intensity at this polarization angle (Supplementary Figure 3). For the higher frequency range (>70 cm$^{-1}$), the spectra for 45° polarization are shown because some closely spaced peaks (P9 and P10 or P11 and P12) are best resolved at this polarization angle. (See Supplementary Figures 4–6 for spectra in different polarizations.) The low-frequency region of the spectrum shows the most pronounced thickness dependence. Both shear and breathing modes are observed. The breathing mode is stronger for the 632.8-nm excitation, but the shear mode is only slightly stronger for the 514.5-nm excitation. Both modes have the same polarization dependence (Supplementary Figure 3). The sharp peak at ~9 cm$^{-1}$ for bulk WTe$_2$ seems to be an overlap of the breathing mode and a shear mode, but the sharpness of the peak seems to indicate that the contribution from the shear mode is stronger. In the higher frequency range, several peaks show clear shift with thickness. There have been several attempts to correlate the Raman spectrum with the thickness of WTe$_2$ [8,23,26–28]. The difference between the two Raman peaks positioned at ~160 cm$^{-1}$ (P10) and ~210 cm$^{-1}$ (P11 and P12) has been proposed as an indicator of the thickness of WTe$_2$. However, because the Raman peak near ~210 cm$^{-1}$ is in fact a superposition of two separate Raman peaks, P11 and P12, with strong polarization dependence (Supplementary Figure 4), its position is not well defined unless the polarization is specified. The significant scatter of the reported separation between the peak at ~210 cm$^{-1}$ and the peak at ~160 cm$^{-1}$ (Supplementary Figure 7) can be thus explained. We find that the separation between P2 and P3 also depends on the thickness. Figures 3(b) and (c) summarize the positions of P2, P3, P10, P11, and P12 as a function of thickness. Figure 3(d) compares the difference between these peaks also as a function of thickness. It is clear that all three differences, P3–P2, P11–P10, and P12–P10, correlate well with thickness. However, since one has to deconvolute the peak at ~210 cm$^{-1}$ to determine the positions of P11 and P12, their positions are less reliable. A small difference between the data from 632.8-nm and 514.5-nm excitations is a result of this limitation. For this reason, we propose that P3–P2 is a more reliable measure of the thickness, because these peaks are not convolution of multiple peaks. Also, because P3 is very weak for 514.5 nm but strong and clearly resolved for 632.8 nm, we suggest that the excitation wavelength of 632.8 nm is more preferable (Supplementary Figure 5).

*2.3. Anomalous behaviors of some Raman peaks*

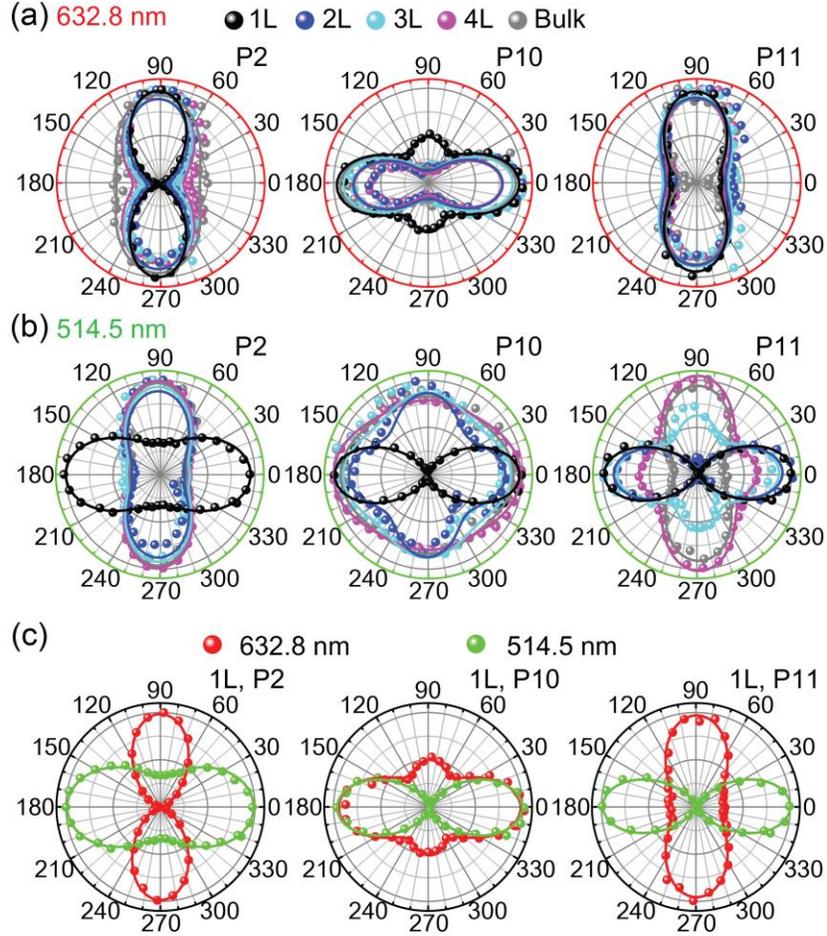

**Figure 4.** Anomalous polarization behaviors of some Raman peaks. Polarization dependence of intensities of peaks P2, P10, and P11 for different thicknesses measured with (a) 632.8-nm and (b) 514.5-nm excitation wavelengths. (c) Polarization dependence of intensities of peaks P2, P10, and P11 for mono-layer WTe$_2$ measured with 632.8-nm and 514.5-nm excitation wavelengths. The points in all polar plots are the normalized Raman intensity data, and the curves are results of fitting to Eq. (2).

Figure 4(a) shows the polarization behaviors of some representative peaks measured with the 632.8-nm excitation. We take the *a*-axis direction (0°) of few-layer WTe$_2$ so that the polarization dependences of the Raman peaks coincide with those of bulk WTe$_2$. The polarization behaviors are essentially identical for all thicknesses, which give credence to the above assignment of the *a*-axis direction. On the other hand, Figure 4(b) shows that for 514.5-nm excitation, there is a strong thickness dependence of the polarization behavior. For example, the polarization behavior of P2 is similar to that for 632.8-nm excitation for all thicknesses except for mono-layer $WTe_2$, which shows an opposite polarization dependence. In the case of P11, the polarization behavior varies gradually from mono-layer to bulk $WTe_2$. (See Supplementary Figures. 6, 8, and 9 for complete data.) Figure 4(c) compares the polarization behaviors of some of the Raman peaks from mono-layer $WTe_2$, measured with excitation wavelengths of 632.8 nm and 514.5 nm.

Similar anomalous polarization behavior that depends on the excitation wavelength or the sample thickness was observed in black phosphorus. The anisotropic electron-phonon interaction as well as the interference effect combined with birefringence and dichroism due to anisotropy could partly explain this anomalous behavior [32–34,36]. Unlike black phosphorus, $WTe_2$ has only weak birefringence (Supplementary Figure 10), and so its contribution to the anomalous polarization behavior, if any, would not be large. We propose that dichroism is the major factor that contributes to this anomalous behavior. If we compare the polarization dependence of the Raman intensities with Eq. (2), we find that the variation of the polarization behaviors with the thickness or the excitation wavelength can be accounted for by varying the phase factor, $\phi_{ab} = \phi_a - \phi_b$. The phase factors in the Raman tensor elements are related to the imaginary part of the dielectric function, which is directly correlated with optical absorption. Because the band structure determines the optical absorption through the joint density of states, the phase factors reflect the differences in the band structure. The band structure of $WTe_2$ depends sensitively on the number of layers [4,22], the observed thickness dependence can be explained in terms of the thickness-dependent dielectric function. The excitation wavelength dependence is understood in the same way because the anisotropy in the dielectric function would depend on the wavelength for a given thickness. Supplementary Figures 8 and 9 summarize the polarization dependence of all the peaks. The $A_2$ modes do not show any dependence on the excitation wavelength or the thickness because the phase factor does not affect the intensity in Eq. (3). The polarization behavior of each $A_1$ mode except P5 does not vary much with thickness for the 632.8-nm excitation but shows significant variation for the 514.5-nm excitation. This is somewhat different from the case of black phosphorus, in which a shorter-wavelength excitation showed less variation with thickness [33]. An exception is P9, which appears to be an $A_2$ mode for 632.8 nm but an $A_1$ mode for 514.5 nm. Because P9 is relatively weak for the 514.5-nm excitation, its deconvolution from P10 is less reliable than for the 632.8-nm excitation. We tentatively assign P9 as an $A_2$ mode which is also consistent with theoretical calculations [2,21,26,28].

## 3. Conclusions

By comparing TED measurements with polarization of the Raman spectrum, we found a reliable method to determine the thickness and crystallographic orientation of few-layer $WTe_2$. The polarization dependence of $A_1$ modes, P2, P10, and P11, measured with the 632.8-nm excitation can be used to determine the crystallographic orientation of 1 to 4-layer and bulk $WTe_2$. On the other hand, the polarization dependence changes with thickness when the 514.5 nm excitation is used. We also propose to use the position difference between P3 and P2, measured with the 632.8-nm excitation, for determination of the thickness, because P3 is enhanced for this excitation wavelength. The difference in the polarization dependence of the Raman peak intensities for different excitation wavelengths is explained in terms of energy-dependent optical dichroism due to the anisotropic structure of $WTe_2$.

## 4. Methods

### 4.1. Sample preparation and Raman measurements

$WTe_2$ samples were mechanically exfoliated on p-type silicon substrates covered with 90-nm thick silicon dioxide. Because $WTe_2$ can get degraded in air [8,21,23] (Supplementary Figure 11), we placed the samples in an optical chamber with ultrapure $N_2$ environment immediately after exfoliation. After suitable thin WTe$_2$ samples were identified using an optical microscope, the substrate with WTe$_2$ samples was transferred to a micro vacuum chamber for Raman measurements. All Raman measurements were carried out in a vacuum of 10$^{-5}$ Torr in order to minimize degradation. A long-working-distance objective (40×, N.A. 0.6) with a correction ring was used to focus the laser beam of 514.5-nm and 632.8-nm wavelengths to a spot size of about 1 μm. The Raman signal was dispersed by a spectrometer with a focal length of 550 mm with a grating with 2400 grooves/mm and detected using a charge-coupled device (CCD). For low frequency Raman measurements in the range below 100 cm$^{-1}$, volume holographic filters were used to clean the laser lines and to eliminate the Rayleigh-scattered light (Supplementary Figure 12). A half-wave plate and a polarizer were used to control the polarization of the excitation laser and another half-wave plate was used to compensate for the polarization dependent throughput of the spectrometer (Supplementary Figure 12). Raman scattered light with the same polarization as the excitation laser (parallel polarization) was measured. The laser power was kept at 300 μW to avoid laser-induced damages to the sample. The slight day-to-day fluctuations in the optical alignment, etc., are corrected for by normalizing against the intensity of the Raman peak from a piece of Si, by measuring the same piece in the same polarization. The number of layers of WTe$_2$ was determined by comparing the optical contrast and atomic force microscopy (AFM) measurements after all the Raman measurements were completed (Supplementary Figure 13). The inter-layer spacing of bulk WTe$_2$ is about 0.7 nm, but the height of monolayer WTe$_2$ measured by AFM is about 1 nm, similar to other TMD materials.

*4.2. Transmission electron diffraction (TED) measurements*

For TED measurements, exfoliated WTe$_2$ flakes on the SiO$_2$/Si substrate were transferred onto a TEM grid by a wet direct transfer method. In detail, for comparison between Raman and TED data, TED analysis should be performed at the particular flakes characterized by Raman spectroscopy. The wet direct transfer method without polymer coating is adopted for deterministic transfer because this method is capable of fixing the grid on a particular flake through optical microscopy. To bond WTe$_2$ flake onto the amorphous carbon film on TEM grids, isopropyl alcohol (IPA) was dropped on top of the grid. As IPA evaporates, surface tension induces tight bonding between amorphous carbon and the WTe$_2$ flake [37]. Then, for detaching the grid bonded with the flake from the SiO$_2$/Si substrate, a KOH solution was used for etching away SiO$_2$. In order to minimize damages on WTe$_2$, we soaked the grid/flake/SiO$_2$/Si stack in the KOH solution to etch the surface of SiO$_2$ rapidly. After a few seconds, the grid with the WTe$_2$ flake was detached from the SiO$_2$/Si substrate. Rinsing the KOH residue was the final stage for the preparation of TEM specimens. Electron diffraction pattern analysis was carried out using a transmission electron microscope, FEI Titan$^3$ G2 60-300, operated at 80 kV acceleration voltage.

**Acknowledgements**

This work was supported by a grant from the Center for Advanced Soft Electronics under the Global Frontier Research Program of the MSIP (No. 2011-0031630) and the National Research Foundation of Korea grants funded by the Korea government (MSIP) (No. 2015R1A2A2A01006992).